\documentclass[preprint,nofootinbib, superscriptaddress, amsmath,amssymb,aps, prd]{revtex4}

\usepackage{graphicx}
\usepackage{dcolumn}
\usepackage{bm}

\usepackage{amsmath}
\usepackage{bm}
\usepackage{slashed}
\usepackage{epsfig}
\usepackage{amsfonts}
\usepackage{epstopdf}
\usepackage{color}
\usepackage{float}
\usepackage{makecell}

\begin{document}

\preprint{}

\title{\boldmath Next-to-leading-order QCD corrections to double prompt $J/\psi$ hadroproduction \unboldmath}

\author{Zhi-Guo He}
\affiliation{Department of Physics and Electronics, School of Mathematics and Physics, Beijing University of Chemical Technology, Beijing 100029, China}
\affiliation{{II.} Institut f\"ur Theoretische Physik, Universit\"at Hamburg,
Luruper Chaussee 149, 22761 Hamburg, Germany}
\author{Xiao-Bo Jin}
\affiliation{Center of Advanced Quantum Studies, Department of Physics, Beijing Normal University, Beijing 100875, China}
\author{Bernd A. Kniehl}
\affiliation{{II.} Institut f\"ur Theoretische Physik, Universit\"at Hamburg,
Luruper Chaussee 149, 22761 Hamburg, Germany}

\date{\today}

\begin{abstract}
Within the framework of nonrelativistic-QCD (NRQCD) factorization, we perform a comprehensive investigation of the color singlet (CS) contribution to prompt $J/\psi$ pair production at the CERN Large Hadron Collider at next-to-leading order (NLO) in $\alpha_s$.
Specifically, we compare our NLO predictions with measurements from the LHCb, CMS, and ATLAS Collaborations.
We find that the CS contribution itself can well describe the LHCb data in most of the experimental bins, except in the regions where the fixed-order calculation is spoiled by the emission of soft or hard-collinear gluons.
In the CMS and ATLAS cases, however, the CS predictions greatly undershoot both the measured total and differential cross sections, despite sizable $K$ factors, of 2--3, for the total cross sections.
\end{abstract} 

\maketitle


\section{Introduction}\label{sec:intro}  

The mechanism of single heavy-quarkonium production is still mysterious after decades of efforts from both the theoretical and experimental sides.
To solve this puzzle, it was proposed to further study the prompt production of heavy-quarkonium pairs in hadron collisions~\cite{Barger:1995vx,Qiao:2002rh,Ko:2010xy}.
In such processes, the binding of heavy-quark-antiquark ($Q\overline{Q}$) pairs takes place twice, yielding increased sensitivity to nonperturbative parameters in theoretical models.
Besides the conventional single parton scattering (SPS) mechanism, heavy-quarkonium pairs can also be produced through double parton scattering (DPS) \cite{Kom:2011bd}.
Therefore, the prompt hadroproduction of heavy-quarkonium pairs also provides an excellent laboratory to study the DPS mechanism and to extract its key parameter $\sigma_{\mathrm{eff}}$, which characterizes the parton correlations within the proton.
Moreover, new structures were observed near the $J/\psi$ pair threshold 
\cite{LHCb:2020bwg,CMS:2023owd,ATLAS:2023bft}.
All of this makes the prompt hadroproduction of heavy-quarkonium pairs, especially of $J/\psi$ pairs, a tantalizing topic. 

Experimentally, the observation of double $J/\psi$ hadroproduction can be traced back to experiments with the NA3 spectrometer at the CERN Super Proton Synchrotron \cite{NA3:1982qlq,NA3:1985rmd} in the early 1980s.
In recent years, comprehensive analyses have been carried out by the LHCb~\cite{LHCb:2011kri,LHCb:2016wuo,LHCb:2023ybt}, CMS~\cite{CMS:2014cmt}, ATLAS~\cite{ATLAS:2016ydt}, and ALICE~\cite{ALICE:2023lsn} Collaborations at the CERN Large Hadron Collider (LHC) and the D0 Collaboration~\cite{D0:2014vql} at the Fermilab Tevatron.
They not only reported total cross sections, but also differential cross sections with respect to various single or double $J/\psi$ kinematic variables.
In particular, the latest LHCb measurements include eight types of differential cross sections~\cite{LHCb:2023ybt} enabling very detailed comparisons with theoretical predictions.
Exploiting that the SPS contribution is negligible in the large-(pseudo)rapidity region, the LHCb, ATLAS, and D0 Collaborations successfully separated SPS and DPS contributions and so obtained the results 
$\sigma_{\mathrm{eff}}=11.3\pm1.8~(\mathrm{stat})\pm2.3~(\mathrm{syst})~\mathrm{mb}$~\cite{LHCb:2023ybt}, $\sigma_{\mathrm{eff}}=6.3\pm1.6\pm1.0~\mathrm{mb}$~\cite{ATLAS:2016ydt}, and $\sigma_{\mathrm{eff}}=4.8\pm0.5\pm2.5~\mathrm{mb}$~\cite{D0:2014vql}, which are inconsistent, however.

On the theoretical side, nonrelativistic-QCD (NRQCD) factorization~\cite{Bodwin:1994jh} is the mainstream approach to describe the heavy-quarkonium production mechanism, and remarkable progress has recently been achieved in double prompt $J/\psi$ hadroproduction.
In the latter case, the interplay between the various $Q\overline{Q}$ Fock states significantly varies over kinematic regions, in contrast to single $J/\psi$ production.
Barger et al.~\cite{Barger:1995vx,Ko:2010xy} first pointed out that the fragmentation contribution from the color octet (CO) channel $gg\to 2c\bar{c}({}^3S_1^{[8]})$ should be predominant in the region of large $J/\psi$ transverse momentum $p^{J/\psi}_T$ so providing a powerful indicator for the CO mechanism, being a key feature of NRQCD factorization.
On the other hand, in the low- to moderate-$p^{J/\psi}_T$ region, the color singlet (CS) channel $gg\to 2c\bar{c}({}^3S_1^{[1]})$ is expected to be more important \cite{Qiao:2002rh,Li:2009ug}.
A complete leading-order (LO) analysis revealed that, in the regions of large $J/\psi$ pair invariant mass $m^{\psi\psi}$ and rapidity difference $|\Delta y^{\psi\psi}|$, the overwhelming bulk of cross section stems from channels due to Feynman diagrams involving $t$-channel gluon exchange \cite{He:2015qya,He:2016idc}.
To explain the experimental measurements, LO predictions are far from sufficient.
It was found that higher-order relativistic corrections can considerably reduce the overestimation by LO predictions in the threshold region~\cite{Martynenko:2012tf,Li:2013csa,He:2024ugx,He:2024lrb} and that next-to-leading-order (NLO) QCD corrections can change the shape of the $p_T^{J/\psi}$ distribution dramatically for the 7~TeV LHCb setup~\cite{Lansberg:2013qka,Sun:2014gca} and largely enhance the LO predictions of total cross section for the ATLAS \cite{Sun:2023exa} and CMS \cite{Sun:2023exacms} setups.
To resolve the inconsistency in the $\sigma_{\mathrm{eff}}$ determination across experiments, large logarithms of the type $(\alpha_s\ln|t/s|)^n$ were resummed by utilizing the parton Reggeization approach (PRA) \cite{He:2019qqr} and partial loop-induced contribution of order $\alpha_s^6$ were considered \cite{Lansberg:2019fgm}.
More theoretical progress on double prompt $J/\psi$ hadroproduction is reported in Ref.~\cite{He:2021oyy} and references cited therein.

The projection of free $Q\bar{Q}$ pairs onto NRQCD Fock states renders computations of higher-order quantum corrections in NRQCD factorization particularly complicated.
It is needless to say that there are new types of infrared divergences in double $P$-wave Fock state production that break NRQCD factorization~\cite{He:2018hwb}.
As for single $J/\psi$ production, almost all the channels were independently evaluated at NLO by more than one group.
To ensure the correctness of the NLO predictions for double $J/\psi$ hadroproduction \cite{Sun:2014gca}, we perform here an independent NLO calculation of the CS contribution to double direct $J/\psi$ hadroproduction.
We compare our results with Ref.~\cite{Sun:2014gca} and find agreement upon adjusting inputs and eliminating a minor inconsistency in the implementation of Ref.~\cite{Sun:2014gca}.
Apart from that, the main purpose of our paper is to present up-to-date comparisons with the latest LHC results \cite{LHCb:2011kri,LHCb:2023ybt,CMS:2014cmt,ATLAS:2016ydt}.

We organize the remainder of this paper as follows.
In Sec.~\ref{sec:calculation}, we describe the theoretical framework for our NLO NRQCD calculation.
In Sec.~\ref{sec:ph}, we present an up-to-date phenomenological discussion of the LHCb \cite{LHCb:2011kri,LHCb:2023ybt}, CMS \cite{CMS:2014cmt}, and ATLAS \cite{ATLAS:2016ydt} measurements, offer a detailed comparison with Ref.~\cite{Sun:2014gca}, and review other known effects, including the NLO relativistic corrections in the CS model \cite{He:2024ugx}, the full set of CO processes at LO in the collinear parton model \cite{He:2015qya} and the PRA \cite{He:2019qqr}, and the BFKL improvement in the latter case \cite{He:2019qqr}.
Our conclusions are summarized in Sec.~\ref{sec:conclusions}.

\section{Theoretical framework}\label{sec:calculation}

Working in the collinear parton model of QCD and in NRQCD factorization, the SPS contribution to the cross section of double prompt $J/\psi$ hadroproduction can be expressed as 
\begin{eqnarray}
\lefteqn{\mathrm{d}\sigma(A+B\to 2J/\psi+X) =
   \sum_{i,j,m,n,H_1,H_2}\mathrm{Br}(H_1 \rightarrow J/\psi+X)
   \mathrm{Br}(H_2 \rightarrow J/\psi+X)}
   \nonumber\\
   &\times&\int\mathrm{d}x_1\mathrm{d}x_2\, f_{i/A}(x_1)f_{j/B}(x_2)
  \mathrm{d}\hat{\sigma}(i+j\to c\bar{c}(m)+c\bar{c}(n)+X)
  \langle\mathcal{O}^{H_1}(m)\rangle\langle\mathcal{O}^{H_2}(n)\rangle,
\end{eqnarray}
where $f_{i/A}(x)$ is the parton distribution function (PDF) of parton $i$ within hadron $A$, $\mathrm{d}\hat{\sigma}$ is the partonic cross section calculated using the standard covariant-projection method \cite{Bodwin:2002cfe}, $\mathrm{Br}(H \rightarrow J/\psi+X)$ is the branching fraction of $H$ decay into $J/\psi$, being 1 for direct production of $H=J/\psi$, and $\langle\mathcal{O}^{H}(n)\rangle$ is the NRQCD long-distance matrix element (LDME) characterizing the formation of heavy quarkonium $H$ from the $Q\overline{Q}$ Fock state $n={}^{2S+1}L_J$ in spectroscopic notation.
We concentrate on the CS state $m=n={}^3S_1^{[1]}$ and the gluon fusion subprocess with $i=j=g$, while we may safely neglect the $q\bar{q}$ annihilation subprocesses, which is greatly suppressed by light-quark PDFs. 

The LO calculation is straightforward.
The NLO corrections include $2\to2$ virtual and $2\to3$ real corrections.
We generate the contributing Feynman diagrams using the program package QGRAF~\cite{Nogueira:1991ex} and handle the Dirac and SU(3)${}_c$ color algebras using FORM~\cite{Vermaseren:2000nd}.
In the loop calculations, we express the form factors of the Feynman amplitudes 
in terms of scalar loop integrals multiplied by coefficients that only depend on Mandelstam variables and the charm quark mass $m_c$.
We employ a custom-made Mathematica code to perform partial fractioning, so as to decompose the denominators of the loop integrals into linearly independent ones.
As cross checks, we also employ the program packages Reduze~2 \cite{vonManteuffel:2012np} and FIRE6 \cite{Smirnov:2019qkx} to reduce the one-loop scalar integrals to a small set of master integrals. 
The master integrals can be found in analytic form with the help of the program packages Package-X 2.0 \cite{Patel:2016fam} and QCDloop \cite{Ellis:2007qk}.
Finally, we use the Cuba library \cite{Hahn:2004fe,Hahn:2014fua} to perform the phase space integrations numerically.

We extract the ultraviolet (UV) and infrared (IR) divergences using dimensional regularization.
We eliminate the UV divergences by renormalizing the parameters and the wave functions of the external legs as usual.
Specifically, we renormalize the charm quark mass $m_c$ and the charm quark and gluon wave functions in the on-mass-shell scheme and the strong coupling $g_s$ in the modified minimal-subtraction ($\overline{\mathrm{MS}}$) scheme.
The counterterms may be found, e.g., in Eq.~(9) of Ref.~\cite{He:2024lrb}, and we refrain from listing them here.

\begin{figure}[t]
\begin{tabular}{cc}
  \includegraphics[width=0.45\textwidth]{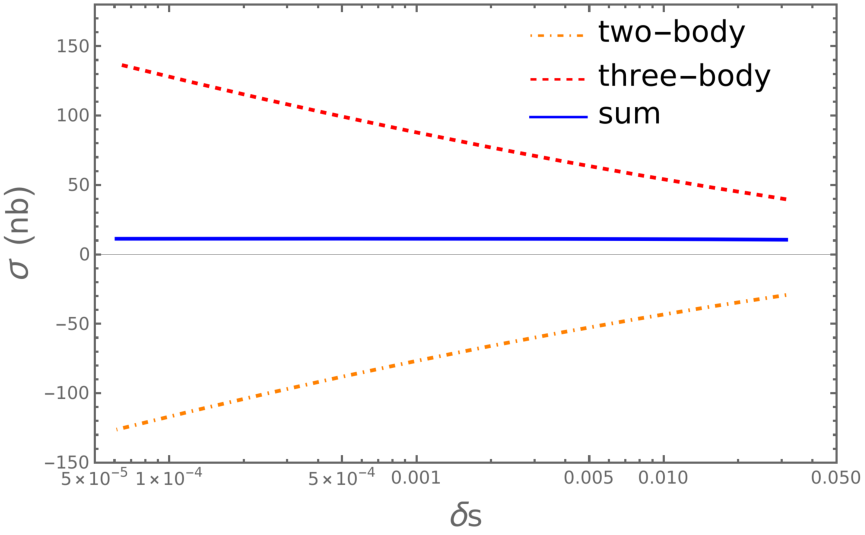}&
  \includegraphics[width=0.45\textwidth]{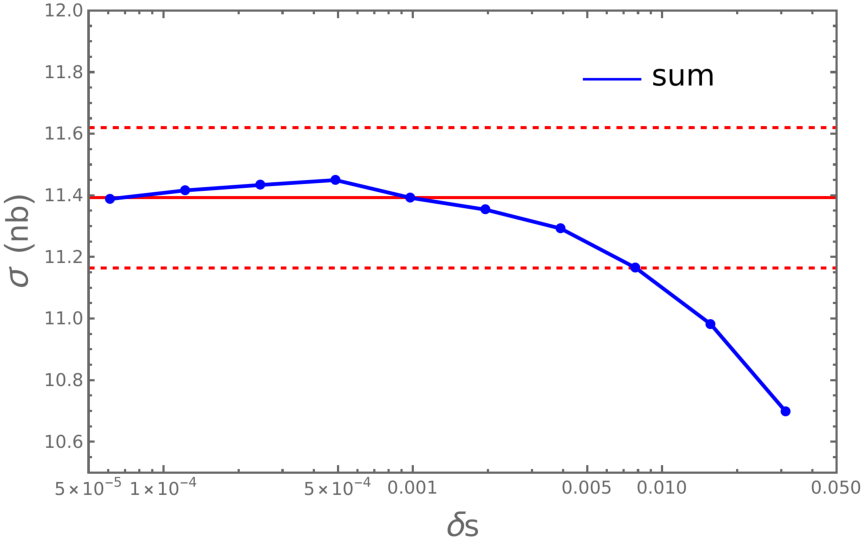}\\
  (a) & (b) 
\end{tabular}
\caption{Dependence of the total cross section for the LHCb setup with $\sqrt{S}=13$~TeV~\cite{LHCb:2023ybt} on the slicing parameter $\delta_s$, with $\delta_c=\delta_s/300$.
(a) The two-body contribution including tree-level, virtual, soft, and hard collinear contributions (orange dot-dashed line) and the three-body contribution made up by the hard non-collinear contribution (red dashed line) sum up to the complete NLO cross section (blue solid line).
(b) The complete NLO cross section is blown up to show its residual variation with $\delta_s$.
The variation in the range $2^{-14}<\delta_s<2^{-7}$ is contained within the $\pm2\%$ error band (red dashed lines) around the central value (red solid line) for $\delta_s=2^{-10}$.
\label{fig:test deltas}}
\end{figure}

The IR divergences from the virtual corrections are mostly canceled by those from the real corrections.
The latter can be divided into soft and collinear parts, which may overlap.
The real corrections arise from the following two $2\to3$ partonic subprocesses:
\begin{eqnarray}
  g+g&\to& (c\bar{c})_1({}^3S_1^{[1]})+(c\bar{c})_2({}^3S_1^{[1]})+g\,,
  \label{eq:gg}\\ 
  g+q(\bar{q})&\to& (c\bar{c})_1({}^3S_1^{[1]})+(c\bar{c})_2({}^3S_1^{[1]})+q(\bar{q})\,.
  \label{eq:gq}
\end{eqnarray}
Partonic subprocess~\eqref{eq:gg} produces both soft and collinear divergences, while partonic subprocess~\eqref{eq:gq} produces only hard collinear divergences.
All of them can be isolated analytically with the help of the two-cutoff phase space slicing method~\cite{Harris:2001sx}.
After combining virtual and real corrections, one is left with IR singularities associated with parton radiation from the incoming legs, which are process independent and may be absorbed into the bare PDFs by the factorization theorem, the resulting PDFs corresponding to the usual $\overline{\mathrm{MS}}$ definition.
The complete IR cancellation serves as a welcome check for the correctness of our NLO calculation.

According to the two-cutoff phase space slicing method \cite{Harris:2001sx}, the three-body phase space is partitioned into soft (S), hard collinear (HC), and hard non-collinear ($\overline{\mathrm{HC}}$) regions by two small parameters, the soft cutoff $\delta_s$ and the collinear cutoff $\delta_c$.
The $2\to3$ cross section is thus decomposed as
\begin{eqnarray}
  \sigma^{\mathrm{real}}=\sigma^{S}+\sigma^{\mathrm{HC}}+\sigma^{\overline{\mathrm{HC}}}\,,
  \label{eq:real}
\end{eqnarray}
where $\sigma^{S}$, $\sigma^{\mathrm{HC}}$, and $\sigma^{\overline{\mathrm{HC}}}$ are all finite, but depend on $\delta_s$ and $\delta_c$.
On the one hand, $\sigma^{S}$ and $\sigma^{\mathrm{HC}}$ are analytically calculated in $D=4-2\epsilon$ dimensions, and the resulting poles in $\epsilon$, due to soft and collinear divergences, are $\overline{\mathrm{MS}}$ subtracted, as they cancel against similar terms in the virtual corrections to the $2\to2$ cross section.
On the other hand, $\sigma^{\overline{\mathrm{HC}}}$ is IR finite and can be computed numerically in $D=4$ dimensions.

While the two-cutoff phase space slicing method \cite{Harris:2001sx} is easy to implement, the cancellation of $\delta_s$ and $\delta_c$ in Eq.~\eqref{eq:real} is never perfect.
If $\delta_s$ and $\delta_c$ become large, then the soft and collinear approximations underlying $\sigma^{S}$ and $\sigma^{\mathrm{HC}}$ deteriorate, while the numerical precision of $\sigma^{\overline{\mathrm{HC}}}$ increases, and vice versa.
In practice, the values of $\delta_s$ and $\delta_c$ are adjusted so that the numerical result is stable.
This is illustrated in Fig.~\cite{LHCb:2023ybt} taking as an example the total cross section evaluated for the 13~TeV LHCb setup \cite{LHCb:2023ybt} (see Table~\ref{Tab:kinematic}) using the inputs specified in Sec.~\ref{sec:ph}.
Specifically, we set $\delta_c=\delta_s/300$ and vary $\delta_s$ between $2^{-14}$ and $2^{-5}$.
We observe from Fig.~\ref{fig:test deltas}(a) how the $\delta_s$ dependencies of the two- and three-body contributions cancel in their sum.
Here, it is understood that the two-body contribution includes the virtual corrections in combination with $\sigma^{S}$ and $\sigma^{\mathrm{HC}}$, while the three-body contribution is just $\sigma^{\overline{\mathrm{HC}}}$.
In Fig.~\ref{fig:test deltas}(b), we zoom in on the latter to visualize its mild
$\delta_s$ dependence.
We consider the total to be sufficiently stable for $\delta_s$ between $2^{-14}\approx6\times10^{-5}$ and $2^{-7}\approx8\times10^{-3}$ and assign to it a central value at $\delta_s=2^{-10}$ with a numerical uncertainty of $\pm2\%$.
This motivates the choice $\delta_s=1/1000$, which we henceforth adopt for our numerical analysis.

\begin{table}
  \caption{\label{Tab:kinematic} Kinematic conditions underlying the LHCb \cite{LHCb:2011kri,LHCb:2023ybt}, CMS \cite{CMS:2014cmt} and ATLAS \cite{ATLAS:2016ydt} measurements.}
    \begin{ruledtabular}
	\begin{tabular}{lcc}
		Experiment & $\sqrt{S}$ [TeV] & Acceptance cuts\\
		\hline
            LHCb 7~TeV~\cite{LHCb:2011kri} & 7 & both $J/\psi$: $p_T<10$~GeV, $2<y<4.5$ \\  
            \hline 
            LHCb 13~TeV~\cite{LHCb:2023ybt} & 13 & both $J/\psi$: $p_T<14$~GeV, $2<y<4.5$\\  
            \hline 
            CMS~\cite{CMS:2014cmt} & 7 & both $J/\psi$: 
            $\left\{ \begin{array}{ll}
                p_T>6.5 \textrm{ GeV }\qquad & \textrm{if }|y|<1.2\\
                p_T>6.5-\frac{2}{0.23}(|y|-1.2) \textrm{ GeV }\qquad & \textrm{if }1.2<|y|<1.43\\
                p_T>4.5 \textrm{ GeV }\qquad & \textrm{if }1.43<|y|<2.2\\
            \end{array} \right.$\\  
            \hline
            ATLAS~\cite{ATLAS:2016ydt} & 8 & 
            \makecell[c]{
            both $J/\psi$: $p_T>8.5$~GeV, $|y|<2.1$\\
            lower-$p_T$ $J/\psi$: $\left\{ \begin{array}{ll}
                |y|<1.05 \qquad& \textrm{central}\\
                1.05<|y|<2.1 \qquad& \textrm{forward}\\
            \end{array} \right.$\\ }
	\end{tabular}
  \end{ruledtabular}
\end{table}

\section{Phenomenology}\label{sec:ph}

We first specify the inputs for our numerical analysis.
At LO (NLO), we employ the CTEQ6L1 (CTEQ6M)~\cite{Pumplin:2002vw} PDFs of the proton and evaluate the strong-coupling constant $\alpha_s^{(n_f)}(\mu_r)$ at one (two) loops with fixed number $n_f=4$ of active quark flavors and asymptotic scale parameter $\Lambda_\mathrm{QCD}^{(n_f)}=215~\mathrm{MeV}$ (326~MeV) \cite{Pumplin:2002vw} using the canonical formula \cite{Kniehl:2006bg}.
For definiteness, we take the on-shell mass of the charm quark to be $m_c=1.5~\mathrm{GeV}$.
We set the renormalization and factorization scales to be $\mu_r=\mu_f=\xi\sqrt{16m_c^2+(p^{J/\psi}_T)^2}$ and vary $\xi$ between $1/2$ and $2$ about the default value 1 to estimate the theoretical uncertainties due to the lack of higher-order corrections.
The CS LDME of the $J/\psi$ meson is determined from the wave function at the origin for the Buchm\"{u}ller--Tye potential~\cite{Eichten:1995ch},
$\langle\mathcal{O}^{J/\psi}({}^3S_1^{[1]})\rangle=1.16~\mathrm{GeV}^3$.
This value agrees with the determination in Ref.~\cite{Bodwin:2007fz},
$\langle\mathcal{O}^{J/\psi}({}^3S_1^{[1]})\rangle=1.32{+0.20\atop-0.17}~\mathrm{GeV}^3$ within errors.

\begin{figure}[tp]
  \includegraphics[width=0.45\textwidth]{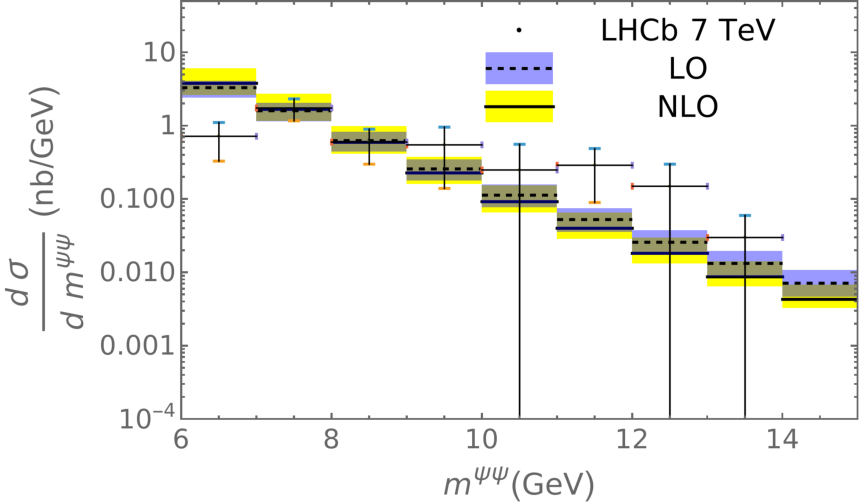}
  \caption{\label{fig:difflhcb7} 
    The LO (dashed lines) and NLO (solid lines) NRQCD predictions for the $m^{\psi\psi}$ distribution $\mathrm{d}\sigma/\mathrm{d}m^{\psi \psi}$ are compared with the LHCb measurements at 7~TeV~\cite{LHCb:2011kri}.
    The theoretical uncertainties at LO and NLO are indicated by the shaded blue and yellow bands, respectively.}
\end{figure}

We are now in the position to compare our theoretical predictions with the measurements by LHCb at center-of-mass energies $\sqrt{S}=7$~TeV \cite{LHCb:2011kri} and 13~TeV \cite{LHCb:2023ybt}, by CMS at 7~TeV \cite{CMS:2014cmt}, and by ATLAS at 8~TeV~\cite{ATLAS:2016ydt}.
The respective kinematic conditions are collected in Table~\ref{Tab:kinematic}.
We begin with the 7~TeV LHCb setup with total cross section 
\begin{eqnarray}
&&\sigma_{\mathrm{LHCb}}=5.1 \pm 1.0 \pm 1.1~\mathrm{nb}\,,
\end{eqnarray}
where the first error is statistical and the second one is systematic.
Our LO and NLO predictions,
\begin{equation}
\sigma^{\mathrm{LO}}_{\mathrm{CS}}=6.02^{+1.56}_{-1.69}{}~\mathrm{nb}\,,\qquad
\sigma^{\mathrm{NLO}}_{\mathrm{CS}}=6.46^{+3.99}_{-2.01}{}~\mathrm{nb}\,,
\end{equation}
both agree with the experimental results within errors.
The NLO corrections are small, the $K$ factor being about 1.07.
From the $m^{\psi\psi}$ distribution shown in Fig.~\ref{fig:difflhcb7}, we observe that the NLO corrections are positive only in the first two bins, while they are negative in the other bins and reach about $-40\%$ in the last bin.
As is familiar from the discussion in Ref.~\cite{He:2024ugx}, the LO CS prediction considerably overshoots the LHCb data near the production threshold at $m^{\psi\psi}\approx6$~GeV.
The NLO corrections somewhat worsen the situation.
The tension is ameliorated by incorporating relativistic corrections, which are negative near threshold \cite{He:2024ugx}.
Except for the first and sixth bins, experiment and theory agree within errors.

\begin{figure}[tp]
\begin{tabular}{ccc}
  \includegraphics[width=0.38\textwidth]{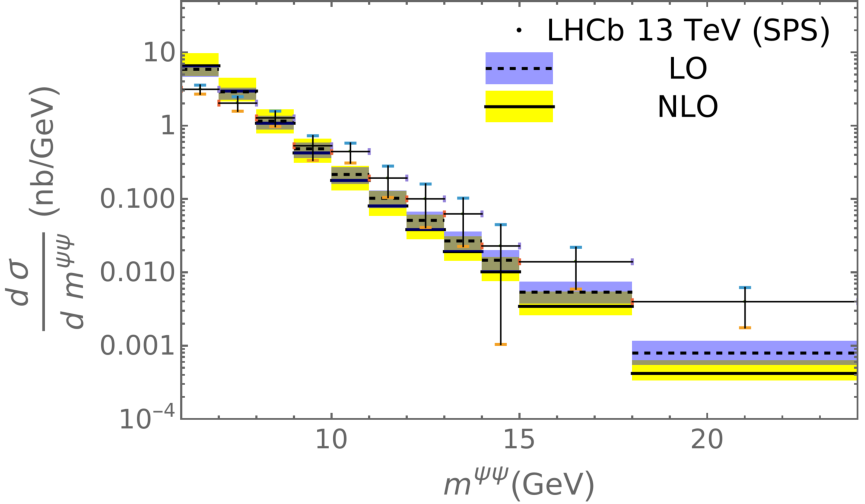}&
  \includegraphics[width=0.38\textwidth]{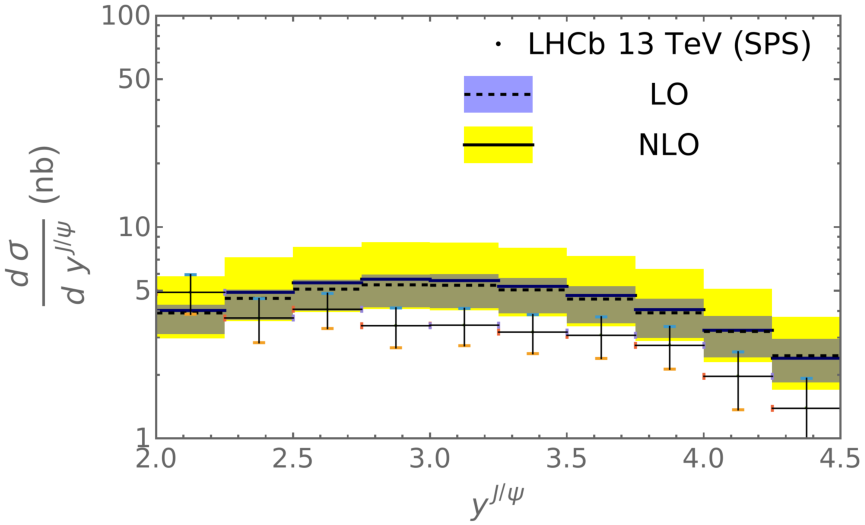}\\
  (a) & (b)\\
  \includegraphics[width=0.38\textwidth]{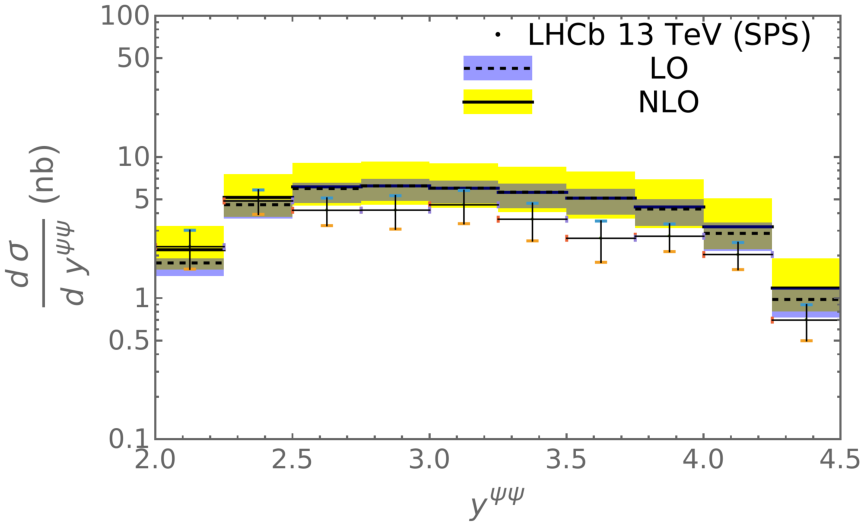}&
  \includegraphics[width=0.38\textwidth]{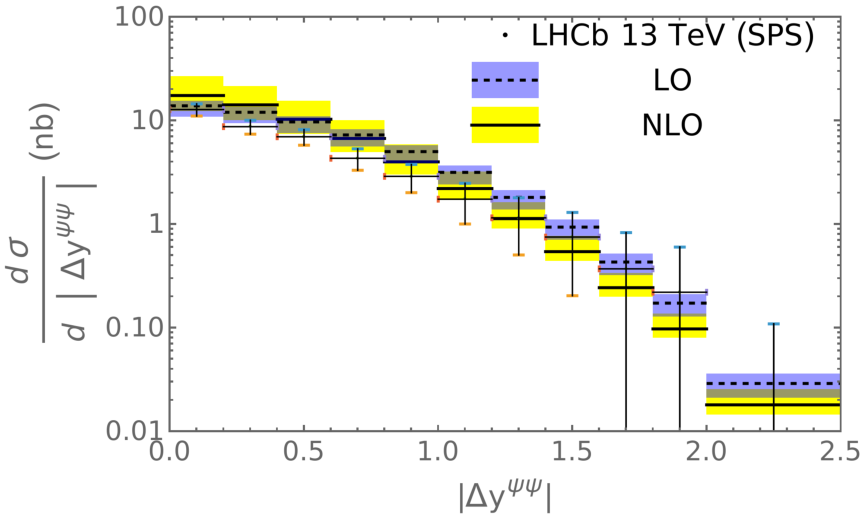}\\
  (c) & (d)\\
  \includegraphics[width=0.38\textwidth]{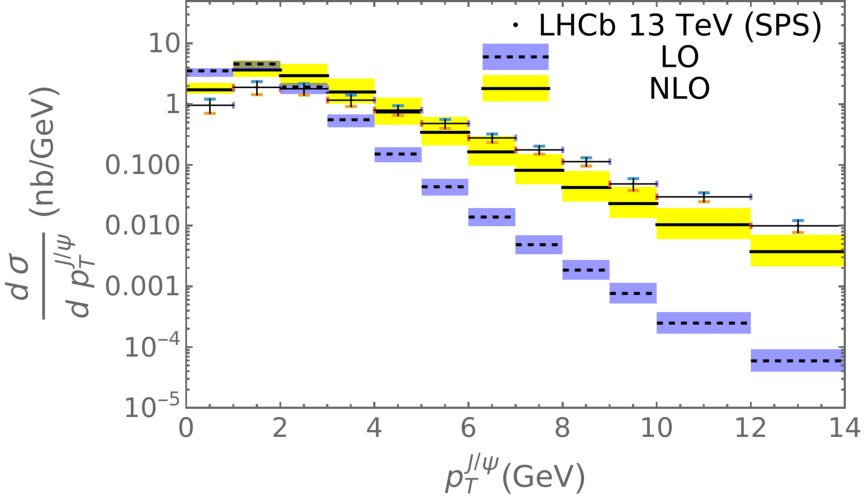}&
  \includegraphics[width=0.38\textwidth]{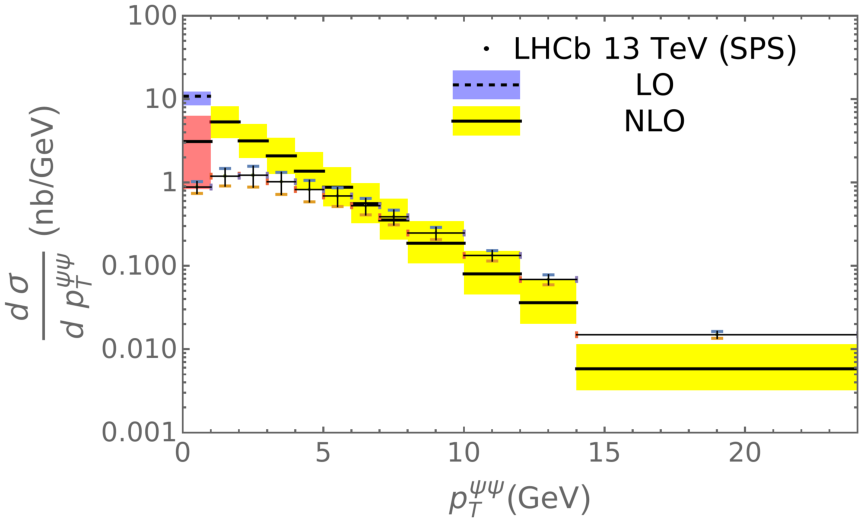}\\
  (e) & (f)\\
  \includegraphics[width=0.38\textwidth]{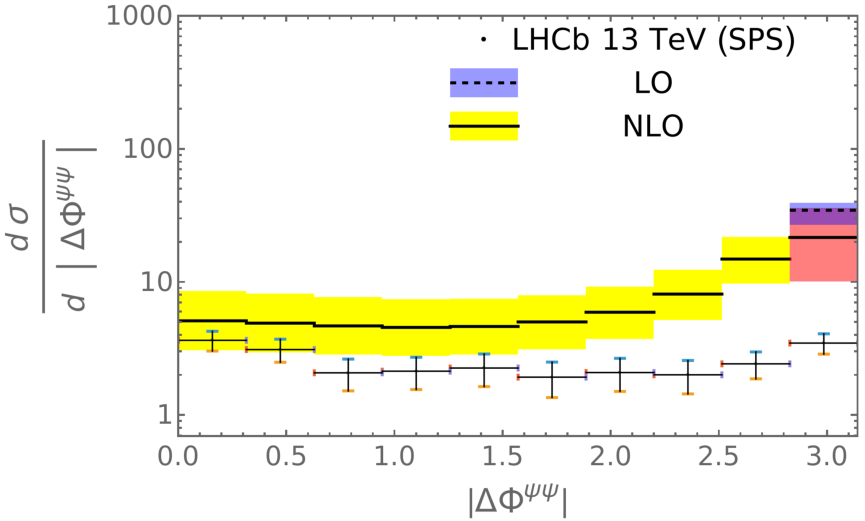}&
  \includegraphics[width=0.38\textwidth]{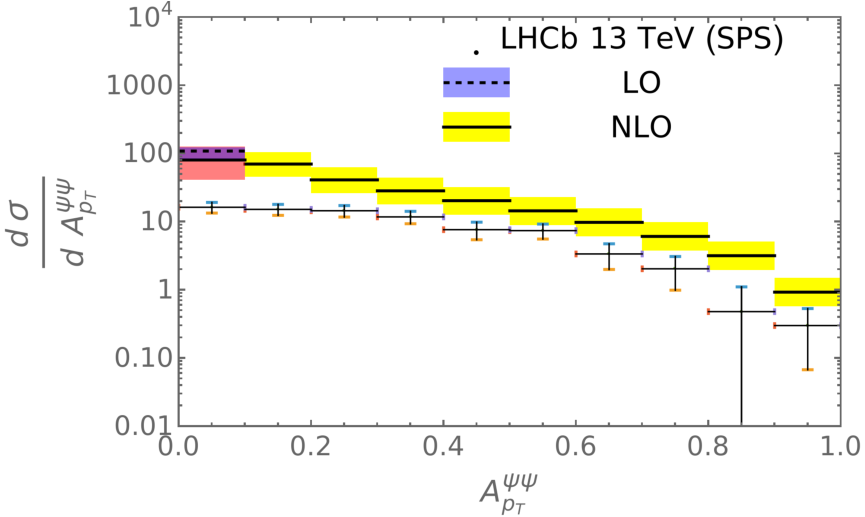}\\
  (g) & (h)\\
\end{tabular}
\caption{\label{fig:difflhcb13}%
The LO (dashed lines) and NLO (solid lines) NRQCD predictions for the differential cross sections
$\mathrm{d}\sigma/\mathrm{d}m^{\psi\psi}$,
$\mathrm{d}\sigma/\mathrm{d}y^{J/\psi}$,
$\mathrm{d}\sigma/\mathrm{d}y^{\psi\psi}$,
$\mathrm{d}\sigma/\mathrm{d}|\Delta y^{\psi\psi}|$,
$\mathrm{d}\sigma/\mathrm{d}p_T^{J/\psi}$,
$\mathrm{d}\sigma/\mathrm{d}p_{T}^{\psi\psi}$,
$\mathrm{d}\sigma/\mathrm{d}|\Delta\Phi^{\psi\psi}|$, and $\mathrm{d}\sigma/dA_{p_T}^{\psi\psi}$ are compared with the respective SPS contributions measured by the LHCb Collaboration at 13~TeV \cite{LHCb:2023ybt}.
The theoretical uncertainties at LO and NLO are indicated by the shaded blue and yellow bands, respectively, except that negative results are indicated by red bands.}
\end{figure}

Next, we move on to the 13~TeV LHCb setup \cite{LHCb:2023ybt}.
Here, the LHCb Collaboration managed to discriminate the SPS contribution from the DPS one, which renders the comparison with our theoretical predictions more meaningful.
For the total cross section of SPS, they quote
\begin{equation}
\sigma_{\mathrm{LHCb}}^{\mathrm{SPS}}=7.9 \pm 1.2 \pm 1.1~\mathrm{nb}\,.
\end{equation}
This agrees within errors with our LO and NLO predictions,
\begin{equation}
  \sigma^{\mathrm{LO}}_{\mathrm{NRQCD}}=10.9^{+1.4}_{-2.5}{}~\mathrm{nb}\,
  \qquad
\sigma^{\mathrm{NLO}}_{\mathrm{NRQCD}}=11.4^{+5.8}_{-3.2}{}~\mathrm{nb}\,,
\end{equation}
although our central values are about 1.4 times larger.
Similarly to the 7~TeV LHCb setup, the $K$ factor is just about 1.05.
From the $m^{\psi\psi}$ distribution shown in Fig.~\ref{fig:difflhcb13}(a), we observe that this overshoot originates from the first bin, near threshold, where higher-order relativistic effects, which are not included here for definiteness, are known to significantly reduce the cross section \cite{He:2024ugx}.
This overshoot near threshold also affects our predictions for other distributions.
On the other hand, in the large-$m^{\psi\psi}$ region, the CS predictions tend to undershoot the experimental data, and this tendency is amplified by the NLO corrections, which reach $-47\%$ in the last bin.
Fortunately, the CO contributions help to fill this gap \cite{He:2015qya,He:2016idc,He:2019qqr,He:2021oyy}.

Besides the $m^{\psi\psi}$ distribution, the LHCb Collaboration measured the cross section at $\sqrt{S}=13$~TeV also differential in $y^{J/\psi}$, $y^{\psi\psi}$, $|\Delta y^{\psi\psi}|=|y^{J/\psi_1}-y^{J/\psi_2}|$, $p_{T}^{J/\psi}$, $p_{T}^{\psi\psi}$, $|\Delta\Phi^{\psi\psi}|=|\Phi^{J/\psi_1}-\Phi^{J/\psi_2}|$ with $\Phi$ being the azimuthal angle, and the transverse-momentum asymmetry,
\begin{equation}
A_{p_T}^{\psi\psi}  = \left| \frac{p_T^{J/\psi_1}-p_T^{J/\psi_2}}{p_T^{J/\psi_1}+p_T^{J/\psi_2}}\right|\,.
\end{equation}
As for the $y^{J/\psi}$, $y^{\psi\psi}$, and $|\Delta y^{\psi\psi}|$ distributions, shown in Figs.~\ref{fig:difflhcb13}(b)--(c), respectively, both our LO and NLO predictions can describe the data.
As in the case of the total cross section, the NLO corrections barely shift the central LO values, except in the region of $|\Delta y^{\psi\psi}|>1$, where the NLO corrections result in a reduction by 30\% to 44\%.

In the case of the $p_{T}^{J/\psi}$ distribution, shown in Fig.~\ref{fig:difflhcb13}(e), the role of the NLO corrections is most prominent.
Here, the LO prediction does not agree with the experimental data at all.
It is about $3.7$ times larger than the experimental data in the first bin and falls short of them by more than two orders of magnitude in the last bin.
This conflict is greatly moderated by the NLO corrections, which reduce the LO central value in the first bin by $51\%$, and boosts it by a factor of 61 in the last bin.

At LO in the collinear parton model, the two $J/\psi$ mesons are produced back to back, so that $p_{T}^{\psi\psi}=A_{p_T}^{\psi\psi}=0$ and $|\Delta \Phi^{\psi\psi}|=\pi$.
Thus, away from these values, in the non-endpoint regions, only the real corrections contribute.
On the other hand, in the small-$p_T^{\psi\psi}$ region, the radiation of soft gluons or hard collinear gluons and quarks are well known to spoil the convergence of perturbation theory.
This is why our NLO predictions are negative in the first bins of $d\sigma/d p_{T}^{\psi\psi}$ and $d\sigma/d|A_{p_T}^{\psi\psi}|$ and in the last bin of $d\sigma/d |\Delta \Phi^{\psi\psi}|$, while they all significantly exceed the experimental data in the neighboring few bins, as may be seen from Figs.~\ref{fig:difflhcb13}(f)--(h).
In the moderate-$p_T^{\psi\psi}$ region, the NLO prediction perfectly agrees with the experimental data.
Yet, as the value of $p_T^{\psi\psi}$ becomes larger, the NLO prediction falls off more strongly than the experimental data, giving room for the CO contributions, which were shown to be non-negligible there using the PRA framework \cite{He:2019qqr,He:2021oyy}.
As for the $d\sigma/d|A_{p_T}^{\psi\psi}|$ and $d\sigma/d |\Delta \Phi^{\psi\psi}|$ distributions, the NLO predictions tend to be consistent with the experimental data away from the endpoint regions.

\begin{figure}[tp]
\begin{tabular}{ccc}
  \includegraphics[width=0.325\textwidth]{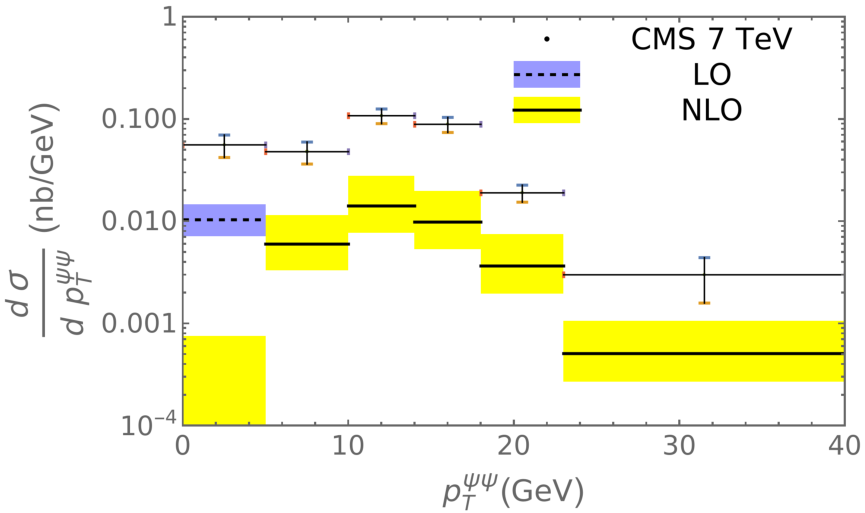}&
  \includegraphics[width=0.325\textwidth]{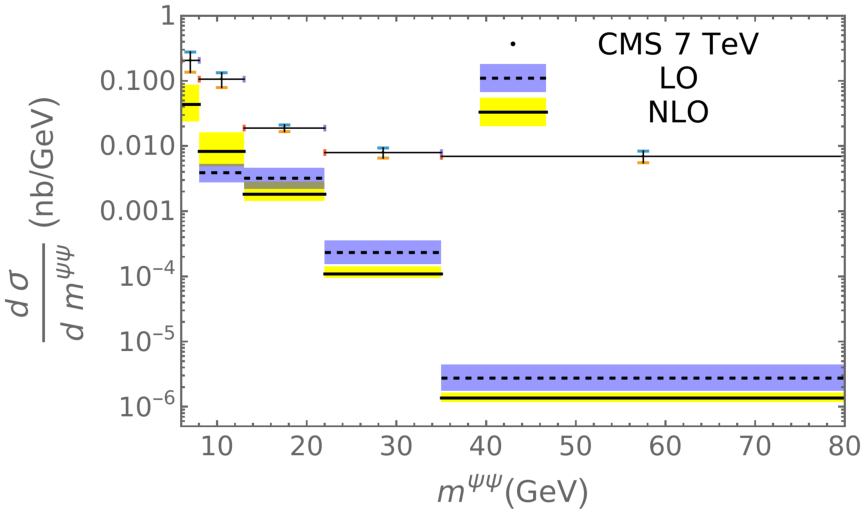}&
  \includegraphics[width=0.325\textwidth]{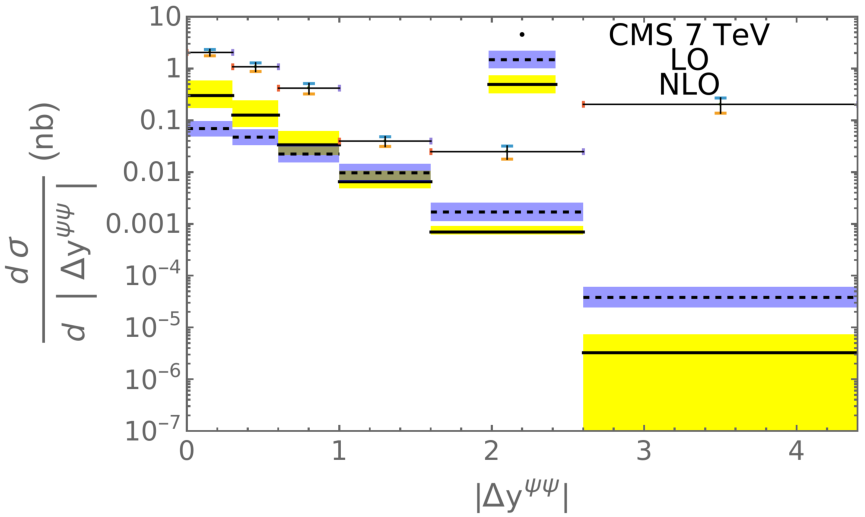}\\
  (a) & (b) & (c)
\end{tabular}
\caption{\label{fig:diffcms}%
  The LO (dashed lines) and NLO (solid lines) NRQCD predictions for the differential cross sections $\mathrm{d}\sigma/\mathrm{d}p_{T}^{\psi\psi}$, $\mathrm{d}\sigma/\mathrm{d}m^{\psi \psi}$, and $\mathrm{d}\sigma/\mathrm{d}|\Delta y^{\psi\psi}|$ 
  are compared with the CMS measurements at 7~TeV~\cite{CMS:2014cmt}.
    The theoretical uncertainties at LO and NLO are indicated by the shaded blue and yellow bands, respectively.}
\end{figure}

We now turn to the CMS measurement at $\sqrt{S}=7$~TeV \cite{CMS:2014cmt}, which imposed a minimum-$p_T^{J/\psi}$ cut at 4.5~GeV and covered a broad rapidity range, $|y^{J/\psi}|<2.1$ (see Table~\ref{Tab:kinematic}).
The total cross section was determined to be
\begin{equation}
\sigma_{\mathrm{CMS}}=1.49 \pm 0.07 \pm 0.13~\mathrm{nb}\,,
\end{equation}
which significantly exceeds our LO and NLO predictions,
\begin{equation}
  \sigma^{\mathrm{LO}}_{\mathrm{NRQCD}}=0.0519^{+0.0214}_{-0.0161}{}~\mathrm{nb}\,,
  \qquad
  \sigma^{\mathrm{NLO}}_{\mathrm{NRQCD}}=0.147^{+0.136}_{-0.061}{}~\mathrm{nb}\,,
\end{equation} 
by factors of 29 and 10, respectively. 
The source of this tremendous gap in total cross section may be understood more deeply by looking at the $p_{T}^{\psi\psi}$, $m^{\psi\psi}$, and $|\Delta y^{\psi\psi}|$ distributions presented in Figs.~\ref{fig:diffcms}(a)--(c).
The NLO prediction for the $p_{T}^{\psi\psi}$ distribution undershoots the experimental data in the whole $p_T^{\psi\psi}$ range by about one order of magnitude, which is quite different from the 13~TeV LHCb case in Fig.~\ref{fig:difflhcb13}(f).
As for the $m^{\psi\psi}$ and $|\Delta y^{\psi\psi}|$ distributions, the LO predictions systematically undershoot the experimental data by one to three orders of magnitude, the undershoot in either case being largest in the last bins.
The NLO corrections ameliorate the situation for $m^{\psi\psi}<13$~GeV and $|\Delta y^{\psi\psi}|<1$, but exacerbate it in the residual bins, where the reductions reach about 50\% and 90\%, respectively.
Fortunately, the CO contributions appreciably enhance the cross section at large values of $m^{\psi\psi}$ and $|\Delta y^{\psi\psi}|$ \cite{He:2015qya,He:2016idc}, especially upon the PRA treatment of initial-state soft-gluon radiation and the Balitsky--Fadin--Kuraev--Lipatov (BFKL) resummation of large logarithmic corrections of the type $(\alpha_s\ln|s/t|)^n$ \cite{He:2019qqr,He:2021oyy}, albeit leaving room for DPS contributions.

\begin{figure}[htp]
\begin{tabular}{ccc}
  \includegraphics[width=0.41\textwidth]{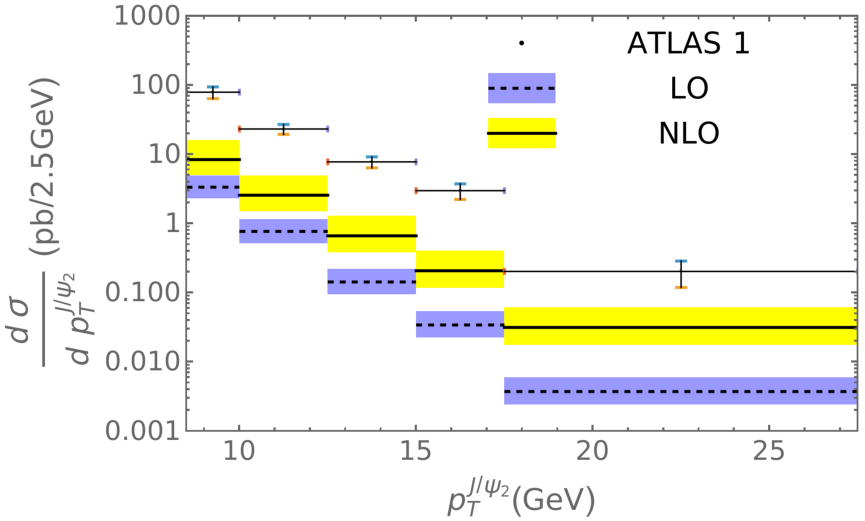}&
  \includegraphics[width=0.41\textwidth]{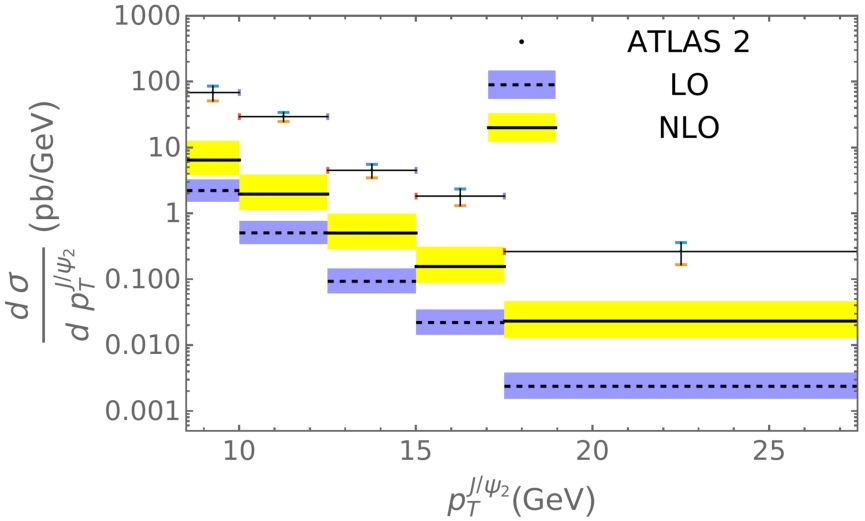}\\
  (a) & (b)\\
  \includegraphics[width=0.41\textwidth]{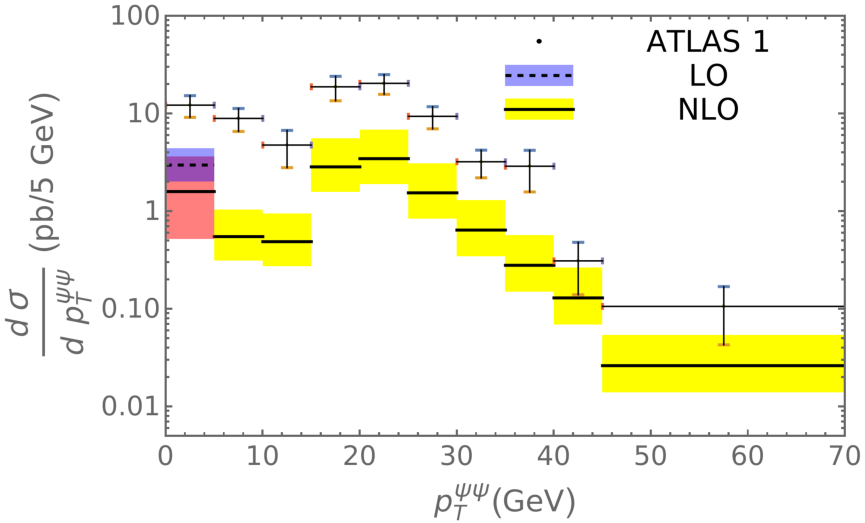}&
  \includegraphics[width=0.41\textwidth]{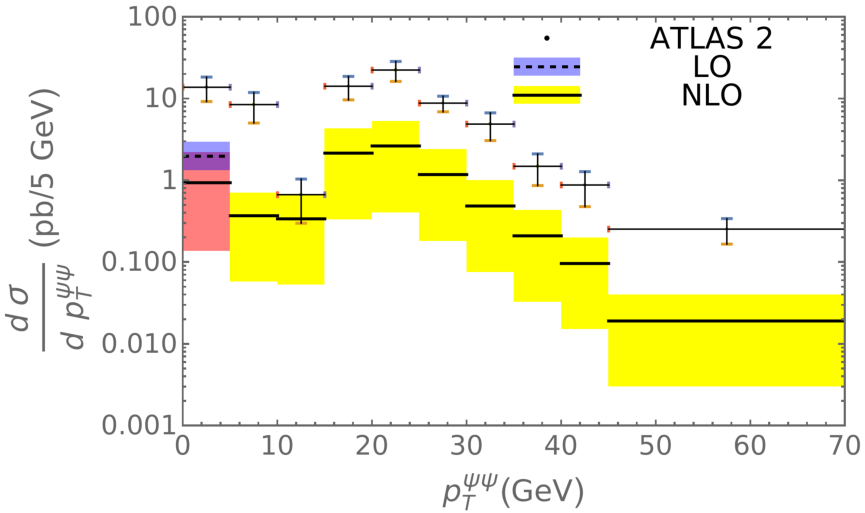}\\
  (c) & (d)\\
  \includegraphics[width=0.41\textwidth]{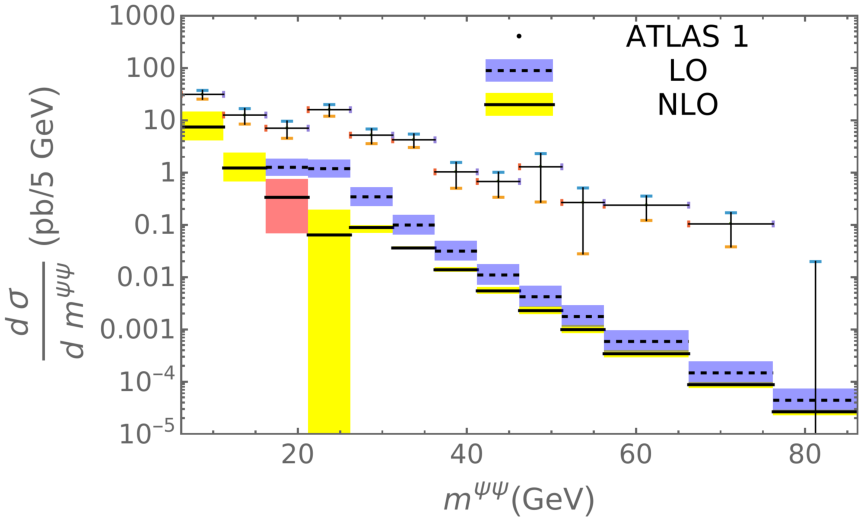}&
  \includegraphics[width=0.41\textwidth]{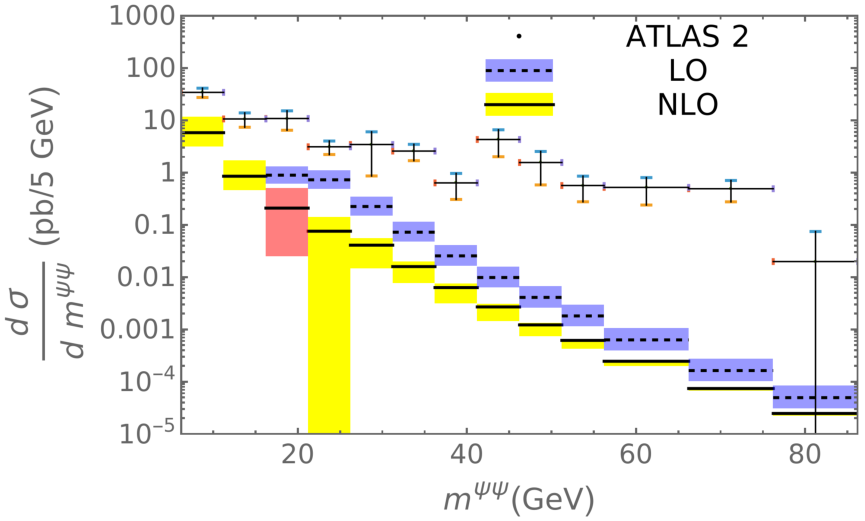}\\
  (e) & (f)\\

\end{tabular}
\caption{\label{fig:diffatlas}%
  The LO (dashed lines) and NLO (solid lines) NRQCD predictions for the differential cross sections $\mathrm{d}\sigma/\mathrm{d}p_T^{J/\psi_2}$, $\mathrm{d}\sigma/\mathrm{d}p_{T}^{\psi\psi}$, and $\mathrm{d}\sigma/\mathrm{d}m^{\psi \psi}$ in the regions of central and forward rapidity $y^{J/\psi_2}$ of the lower-$p_T$ $J/\psi$ meson are compared with the respective ATLAS measurements at 8~TeV~\cite{ATLAS:2016ydt}, labeled ATLAS~1 and ATLAS~2.
The theoretical uncertainties at LO and NLO are indicated by the shaded blue and yellow bands, respectively, except that negative results are indicated by red bands.}
\end{figure}

Finally, we are left with the ATLAS measurement at $\sqrt{S}=8$~TeV \cite{ATLAS:2016ydt}, with an even higher minimum-$p_T^{J/\psi}$ cut, at 8.5~GeV, which comes in two samples depending on whether the $J/\psi$ meson with the lower $p_T$ value ($J/\psi_2$) was detected in the central or forward regions of the detector (see Table~\ref{Tab:kinematic}).
They reported total cross sections of
\begin{equation}\label{eq:atlas}
\sigma_{\mathrm{ATLAS}}=\left\{ \begin{array}{ll}
                82.2 \pm 8.3 \pm 6.3  \pm 0.9 \pm 1.6~\mathrm{pb}\,,\qquad & |y^{J/\psi_2}|<1.05\\
                78.3 \pm 9.2 \pm 6.6 \pm 0.9 \pm 1.5~\mathrm{pb}\,,\qquad & 1.05<|y^{J/\psi_2}|<2.1
            \end{array} \right.
\end{equation}
where the errors are due to statistics, systematics, $J/\psi\to\mu^+\mu^-$ branching fraction, and luminosity, respectively.
The results in Eq.~\eqref{eq:atlas} are estimated to include a modest DPS admixture, of 10\% \cite{ATLAS:2016ydt}.
On the other hand, our LO and NLO predictions are
\begin{eqnarray}
\sigma^{\mathrm{LO}}_{\mathrm{NRQCD}}&=&\left\{ \begin{array}{ll}
            2.97^{+1.42}_{-0.96}{}~\mathrm{pb}\,,\qquad & |y^{J/\psi_2}|<1.05\\
            1.98^{+0.97}_{-0.65}{}~\mathrm{pb}\,,\qquad &  1.05<|y^{J/\psi_2}|<2.1
\end{array} \right.
\nonumber\\
\sigma^{\mathrm{NLO}}_{\mathrm{NRQCD}}&=&\left\{ \begin{array}{ll}
            8.61^{+7.87}_{-3.56}{}~\mathrm{pb}\,,\qquad & |y^{J/\psi_2}|<1.05\\
            6.64^{+6.36}_{-2.82}{}~\mathrm{pb}\,,\qquad &  1.05<|y^{J/\psi_2}|<2.1
            \end{array} \right.
\end{eqnarray}
which are about 28 (40) and 10 (12) times smaller in the central (forward) $y^{J/\psi_2}$ region, respectively.
The substantial $K$ factors, of 2.9 and 3.4, are not nearly large enough to fill up the huge gaps with respect to the experimental data.

We compare our LO and NLO predictions for the $p_{T}^{J/\psi_2}$, $p_{T}^{\psi\psi}$, and $m^{\psi\psi}$ distributions with the experimental data in Figs.~\ref{fig:diffatlas}(a)--(c).
As for the $p_{T}^{J/\psi_2}$ distributions, the LO predictions systematically undershoot the experimental data both in the central and forward $y^{J/\psi_2}$ regions, by up to three orders of magnitude.
These gaps are partly filled by substantial NLO corrections, whose $K$ factors range from $2.5$ and $2.9$ in the first bins to $8.5$ and $9.8$ in the last bins.
As in the CMS case in Fig.~\ref{fig:diffcms}(a), the predicted $p_{T}^{\psi\psi}$ distribution nicely reproduces the shape of the experimental data, but systematically falls short of it.
Incidentally, there is agreement within errors in the last two bins of the central $y^{J/\psi_2}$ region and in the third bin of the forward $y^{J/\psi_2}$ region.
Similarly to Figs.~\ref{fig:difflhcb13}(f)--(h), the negative NLO predictions in the first bins are due to the demarcation between $2\to2$ and $2\to3$ kinematics at fixed order.
As for the $m^{\psi\psi}$ distributions, we note that the LO predictions first appear in the third bin because of the large minimum-$p_T^{J/\psi}$ cut.
By the same token, only the $2\to3$ partonic subprocesses of the real NLO corrections contribute in the first two bins.
This also explains the negative NLO predictions in the third bins, which are in line with Figs.~\ref{fig:difflhcb13}(f)--(h) and Figs.~\ref{fig:diffatlas}(c) and (d).
The NLO predictions undershoot the experimental data in the first two bins by typically one order of magnitude.
The LO predictions significantly undershoot the experimental data and fall off more steeply with $m^{\psi\psi}$ than the latter.
From the fourth bins onward, the NLO corrections are all negative, ranging from $-95\%$ and $-90\%$ in the fourth bins to $-40\%$ and $-50\%$ in the last bins of the central and forward $y^{J/\psi_2}$ regions, respectively.

At this point, we compare our LO and NLO predictions with Ref.~\cite{Sun:2014gca}, adopting all inputs from there.
We do this for the total cross section under 7~TeV LHCb kinematic conditions \cite{LHCb:2011kri}. 
Our LO and NLO results, $4.41^{+1.23}_{-1.25}$~nb and $4.57^{+2.79}_{-1.40}$~nb, agree with theirs, $4.56\pm 1.13$~nb and $5.41^{+2.73}_{-1.14}$~nb, within the theoretical uncertainties.
We observe that their central LO and NLO values exceed ours by 3\% and 18\%, respectively.
Comparing their $K$ factor, 1.19, with ours, 1.04, we observe that their NLO corrections are about five times larger than ours.
This suggests that our NLO calculation disagrees with theirs.
After detailed communications \cite{Sun:2025com} with one of the authors of Ref.~\cite{Sun:2014gca}, we are in a position to explain the differences mentioned above.
For one thing, the authors of Ref.~\cite{Sun:2014gca} evaluate $\alpha_s$ as implemented in Ref.~\cite{Pumplin:2002vw}, while we use the canonical formula \cite{Kniehl:2006bg}.
This explains the slight mismatch at LO and also a part of the difference at NLO.
The residual difference at NLO can be traced to a slightly incorrect implementation of the hard-collinear corrections in the $gq$ and $qg$ channels in Ref.~\cite{Sun:2014gca}.
Evaluating $\alpha_s$ as in Ref.~\cite{Pumplin:2002vw} and implementing the above-mentioned inconsistency at NLO in our computer code, we are able to reproduce the results in Ref.~\cite{Sun:2014gca} within numerical uncertainties.

Starting from the LO prediction in the CS model, we have included here the NLO quantum corrections.
It is interesting to quantitatively compare this improvement with other effects that were considered in the literature, including the NLO relativistic corrections in the CS model \cite{He:2024ugx}, the full set of CO processes at LO in the collinear parton model \cite{He:2015qya} and the PRA \cite{He:2019qqr}, and the BFKL improvement in the latter case \cite{He:2019qqr}.
Specifically, Fig.~\ref{fig:difflhcb7} can be compared with Fig.~2 in Ref.~\cite{He:2015qya}; Figs.~\ref{fig:difflhcb13}(a), (b), and (d) with Figs.~1(a), (c), and (b) in Ref.~\cite{He:2024ugx}; Fig.~\ref{fig:diffcms}(a) with Fig.~2(a) in Ref.~\cite{He:2019qqr}; Figs.~\ref{fig:diffcms}(b) and (c) with Figs.~3 and 4 in Ref.~\cite{He:2015qya} and with Fig.~2(b) and (c) in Ref.~\cite{He:2019qqr}; and Figs.~5(a)--(f) with Figs.~3(a), (d), (b), (e), (c), and (f) in Ref.~\cite{He:2019qqr}.
The relativistic corrections \cite{He:2024ugx} were found to reduce the cross section significantly in the smallest $m^{\psi\psi}$ bin, by $44\%$, and moderately in the residual phase space.
The CO processes \cite{He:2015qya} turned out to greatly dominate the cross section at large values of $m^{\psi\psi}$ and $|\Delta y^{\psi\psi}|$ values.
The treatment in the PRA \cite{He:2019qqr} enabled kinematics that is only amenable at NLO in the collinear parton model, such as finite $p_T^{\psi\psi}$ values, and yielded satisfactory agreement with the experimental data in general, except for the large-$m^{\psi\psi}$ and large-$|\Delta y^{\psi\psi}|$ regions, where the predictions were found to greatly undershoot the experimental data.
In the latter regions, BFKL resummation \cite{He:2019qqr} was seen to enhance the cross section by up to a factor of two and so to improve the description of the experimental data. 

\section{Conclusions}\label{sec:conclusions}

The study of double prompt $J/\psi$ hadroproduction provides a formidable opportunity to deepen our understanding of the $J/\psi$ production mechanism and to shed light on the notorious $J/\psi$ polarization puzzle.
In fact, the contributing LDMEs are weighted there quite differently than in single prompt $J/\psi$ hadroproduction thus providing orthogonal information in global fits to experimental data. 
To obtain reliable theoretical predictions, it is indispensable to go to NLO.
As for fixed-order calculations, only the CS channel had been fully explored at NLO in $\alpha_s$ \cite{Sun:2014gca}, and an independent verification had been missing so far.
This motivated our present analysis, which has reached agreement with the one in Ref.~\cite{Sun:2014gca} upon adjusting inputs and correcting a minor inconsistency in Ref.~\cite{Sun:2014gca}.

As for phenomenology, we found that, in the low-$p_T^{J/\psi}$ region accessed by the 7~TeV \cite{LHCb:2011kri} and 13~TeV \cite{LHCb:2023ybt} LHC measurements, the NLO CS prediction describes the experimental data reasonably well, except in two regions.
One is the threshold region, where higher-order relativistic corrections should be taken into account~\cite{He:2024ugx}.
The other one is the small-$p_T^{\psi\psi}$ region, where resummation techniques are needed, which are, however, beyond the scope of this work.
Furthermore, although the NLO corrections have little influence on the LO predictions of the total cross section and the $y^{J/\psi}$ and $y^{\psi\psi}$ distributions for the 13~TeV LHCb setup \cite{LHCb:2023ybt}, they significantly change the shape of the $p_T^{J/\psi}$ distribution thus improving the agreement with the experimental data, and only the real corrections contribute to the $p_T^{\psi\psi}$, $|\Delta \Phi^{\psi\psi}|$, and $|A_{p_T}^{\psi\psi}|$ distributions in the non-endpoint regions.
In the large-$p_T^{J/\psi}$ region, where the CMS \cite{CMS:2014cmt} and ATLAS \cite{ATLAS:2016ydt} data are settled, the CS contribution becomes minor for both the total and differential cross sections.
This feature is especially pronounced in the large-$m^{\psi\psi}$ and large-$|\Delta y^{\psi\psi}|$ regions.

We conclude from our study that double prompt $J/\psi$ hadroproduction is far from being well understood and requires further investigation.

\subsection*{Acknowledgments}

We thank Li-Ping Sun for the fruitful collaboration in comparing the results of Ref.~\cite{Sun:2014gca} with ours.
X.-B. Jin would like to thank Yan-Qing Ma for the hospitality and helpful discussions during his visit at the School of Physics, Peking University.
This work was supported in part by the German Research Foundation DFG through Grants No.~KN~365/13-2 and 365/14-2, by the National Natural Science Foundation of China through Grants No. 12135013, and by Fundamental Research Funds for the Central Universities through Grant No. buctrc202432.

\end{document}